\documentclass{article}

\usepackage{PRIMEarxiv}

\usepackage[utf8]{inputenc} 
\usepackage[T1]{fontenc}    
\usepackage{hyperref}       
\usepackage{url}            
\usepackage{booktabs}       
\usepackage{amsfonts}       
\usepackage{amsmath}        
\usepackage{amssymb}        
\usepackage{nicefrac}       
\usepackage{natbib}         
\usepackage{microtype}      
\usepackage{lipsum}
\usepackage{fancyhdr}       
\usepackage{graphicx}       
\graphicspath{{media/}}     
\usepackage[version=4]{mhchem}
\usepackage[table]{xcolor}
\usepackage{algorithm}
\usepackage{algorithmic}
\usepackage{multirow}
\usepackage{graphicx}
\usepackage{fancyvrb}
\usepackage{tablefootnote} 

\title{Navigating Order-(Dis)Order Family Trees via \\ Group-Subgroup Transitions
}

\author{
  Shuya Yamazaki \\
  Nanyang Technological University \\
   \And
  Yuyao Huang \\
  Nanyang Technological University \\
  \AND
  Martin Hoffmann Petersen \\
  Nanyang Technological University \\
  \And
  Wei Nong \\
  Nanyang Technological University \\
  \And
  Kedar Hippalgaonkar\thanks{Correspondence to \texttt{kedar@ntu.edu.sg}} \\
  Nanyang Technological University \\
}

\begin{document}
\maketitle

\begin{abstract}
As closed-loop materials discovery systems scale to produce millions of candidate compounds,  the credibility of the novelty they reward becomes a critical concern. Novelty is commonly assessed against databases of ordered crystal structures, in which atomic sites are fully occupied. Yet, a predicted ordered structure may simply correspond to a particular ordering of a known disordered phase, whose sites are occupied by multiple species in the statistical average structure; we refer to such a structure as an ordered child of a disordered parent. Here, we introduce order-(dis)order family trees, a symmetry-based framework that organizes ordered and disordered structures through group-subgroup relations and enables novelty to be explicitly evaluated. We develop a high-throughput family matching procedure, to identify possible disordered parents and symmetry-related ordered relatives for a given ordered structure. As validation, we test our framework on synthesis-facing case studies (A-Lab), where it correctly recovers existing disordered parents for the targeted ordered structures. Extending this family-tree-based benchmark to experimental structure databases (ICSD), computational datasets (MP-20, Alex-MP-20, and GNoME), and crystal generative models further reveals that many ordered structures that appear novel as individual entries are, in fact, better understood as members of experimentally known order-(dis)order family trees. We also show that this is particularly evident in symmetry-agnostic all-atom generative models, which more frequently produce ordered structures derived from known disordered parents, whereas symmetry-constrained models are $2$--$4\times$ less prone to this behavior. Our results establish order-(dis)order family trees as a key requirement for achieving genuine novelty in data-driven materials discovery.
\end{abstract}


\section{Introduction}
Materials discovery systems are pushing the field toward a closed loop between computational prediction and experimental synthesis. Agentic workflows now produce millions of candidate compounds at a rate that far outpaces experimental characterization, as they scale from proposing structures to directing robotic synthesis~\cite{Szymanski2023, Merchant2023}. Consequently, the central question is no longer simply whether we can predict stable structures beyond the training distribution, but whether the structures we predict are genuinely new to synthesis~\cite{Cheetham2024, Leeman2024}.  In current practice, novelty is typically defined by absence from reference databases of ordered crystal structures~\cite{Xie2021, Zeni2025, Betala2025}. Defining novelty in these systems has become a critical concern: a system that rediscovers known phases dressed as new compounds wastes synthesis cycles and corrupts its own reward. 

Yet, answering whether a predicted structure is genuinely novel is more subtle than it first appears. Novelty is not an intrinsic property of a compound, but one defined relative to a reference: a structure is novel only with respect to the space against which it is compared, and the choice of that reference therefore shapes what counts as discovery. Although ordered structure databases are a natural reference given the data on which many discovery systems are trained, relying on them alone introduces a systematic blind spot. Experimentally synthesized materials frequently exhibit occupational disorder, in which the same crystallographic lattice sites are occupied by multiple species in the statistical average structure~\cite{Rhl2019}; hereafter, disordered phase refers to this average crystallographic structure unless otherwise noted. In this setting, a predicted ordered structure may not correspond to a genuinely new compound, but instead to a specific ordering of a known disordered phase. Such a structure corresponds to one of many possible ordered configurations consistent with the same underlying disordered framework. We refer to such predictions as \emph{ordered children} of \emph{disordered parents}~\cite{Cheetham2024, Leeman2024}. When disordered phases are not included in the reference space, these ordered children are routinely labeled as novel, leading to a systematic overestimation of discovery rates relative to experimental reality.

Recent experimental efforts highlight this exact failure mode. In the A-Lab study~\cite{Szymanski2023}, over $65\%$ of the "successful" syntheses of GNoME compounds~\cite{Merchant2023} were later found to lack the predicted cation ordering, and were instead more plausibly interpreted as already-known disordered parent phases~\cite{Leeman2024}. An example is \ce{MgTi2NiO6}: although the predicted structure was ordered, the synthesized product was more consistent with the known disordered ilmenite-type parent phase at the same composition. The same pattern has emerged in generative-model-driven discovery. In MatterGen~\cite{Zeni2025}, the experimentally synthesized product was not the predicted ordered phase but its disordered parent, subsequently identified as a known phase differing only slightly in site occupancy from previously reported structures~\cite{Juelsholt2026}.

These cases are unlikely to be isolated failures. Rather, they point to a structural gap in how computational predictions are interpreted. Ordered and disordered phases are often not unrelated alternatives, but members of the same crystallographic lineage~\cite{Barnighausen1980}. This connection arises because atomic ordering directly governs crystal symmetry: a disordered phase, in which no specific arrangement of species over atomic sites is enforced, preserves the full symmetry of the underlying lattice, whereas the onset of ordering breaks that symmetry and produces a lower-symmetry structure. These symmetry reductions are not arbitrary but are constrained by group-subgroup relations in crystallography~\cite{Barnighausen1980, Ivantchev2000, Han2025}, governing the ordered structures that can descend from a given disordered parent. Multiple ordered structures can descend from the same parent through distinct symmetry-breaking pathways, forming a family of ordered siblings. When these relations are ignored, a structure can then be erroneously counted as a new compound even though it already occupies a position within a known order-disorder family. 

Such relations are rarely available in a systematic, high-throughput form, and most reported examples have been uncovered only through detailed manual analysis of a small number of predicted structures subjected to experimental validation~\cite{Cheetham2024, Leeman2024, Juelsholt2026}. In principle, one could attempt exhaustive mapping by enumerating all symmetry-inequivalent ordered configurations of known disordered phases, for example using methods such as POCC~\cite{Yang2016}, and then matching a query ordered structure against the resulting lookup space. In practice, however, extending such enumeration across the full set of known disordered compounds is computationally prohibitive and unsuitable for large-scale analysis. More fundamentally, this still does not provide a general framework for answering: (i) for a given ordered structure, what disordered parents it may derive from, (ii) what ordered siblings share the same parents, and, most importantly, (iii) whether its apparent novelty reflects a genuinely new family or merely a new member within an existing one.

To address this gap, we develop a symmetry-based framework that organizes ordered and disordered phases through \textbf{group-subgroup transitions} in space-group hierarchy: \textbf{order-(dis)order family trees}. In this view, a disordered parent is not only a possible precursor or competing interpretation of a predicted ordered phase, but also the root of a broader family tree whose branches include symmetry-lowered ordered descendants and related sibling structures. Group-subgroup transitions provide the natural language for navigating these trees, tracing how structural order emerges, splits, and reconnects across different levels of symmetry.

This family-tree perspective changes how computational discovery should be interpreted. Rather than treating each predicted structure as an isolated point, we view it as an element embedded in a manifold of related ordered and disordered phases. What is discovered computationally is therefore often not a standalone structure, but a position within an existing family tree. This distinction is particularly important during synthesis and characterization of predicted compounds, where experimentally realized phases may actually be disordered parents, or coexist with nearby siblings, or appear as partially ordered variants of the same family. In this framework, we develop a high-throughput procedure, built on the Symmetry and Wyckoff-sequence of Ordered and Disordered crystals representation (SWORD) \cite{Huang2026}, for identifying candidate disordered parents and symmetry-related family members of ordered query structures. Using these mappings, we then define disorder-aware family metrics that evaluate a structure set against both ordered and disordered known phases, rather than against ordered references alone. This yields a view of novelty that better reflects experimental reality by exposing cases where predicted ordered phases only appear novel because their known disordered counterpart is absent from the reference set, providing critical feedback to recent novelty assessments \cite{Xie2021,Zeni2025}.

We first validate our framework on the A-Lab cases, demonstrating that it recovers the disordered parent phases underlying a substantial fraction of the synthesis outcomes. We then introduce two new Family Tree-based novelty metrics, $FT_{\mathrm{disorder}}$ and $FT_{\mathrm{order}}$, which are defined respectively as the fraction of ordered compounds connected through order-disorder and order-order family trees to experimentally known disordered and ordered phases. We then apply these metrics across various experimental and computational structure databases (ICSD~\cite{Zagorac2019}, MP-20~\cite{Xie2021}, Alex-MP-20~\cite{Zeni2025}, GNoME~\cite{Merchant2023}) and crystal generative models (DiffCSP~\cite{Jiao2023}, DiffCSP++~\cite{Jiao2024}, FlowMM~\cite{Miller2024}, MiAD~\cite{Okhotin2025}, SymmCD~\cite{Levy2025}, CrystalDiT~\cite{Yi2025}, ADiT~\cite{Joshi2025}, MatterGen~\cite{Zeni2025}, Chemeleon2-RL~\cite{Park2025}, WyFormer~\cite{Kazeev2025}), finding that a consistent and substantial fraction of apparently novel structures are better understood as members of known order-(dis)order family trees. Therefore, we posit that novelty in materials discovery should be assessed not only by structural distinctness, but also by considering crystallographic lineage. Making order-(dis)order family trees explicit helps distinguish genuinely new families from symmetry-lowered rediscoveries of known phases, thereby providing a clearer bridge between computational prediction and experimental realization.

\section{Methods}
\label{sec:methods}

\begin{figure}[t]
    \centering
    \includegraphics[width=\linewidth]{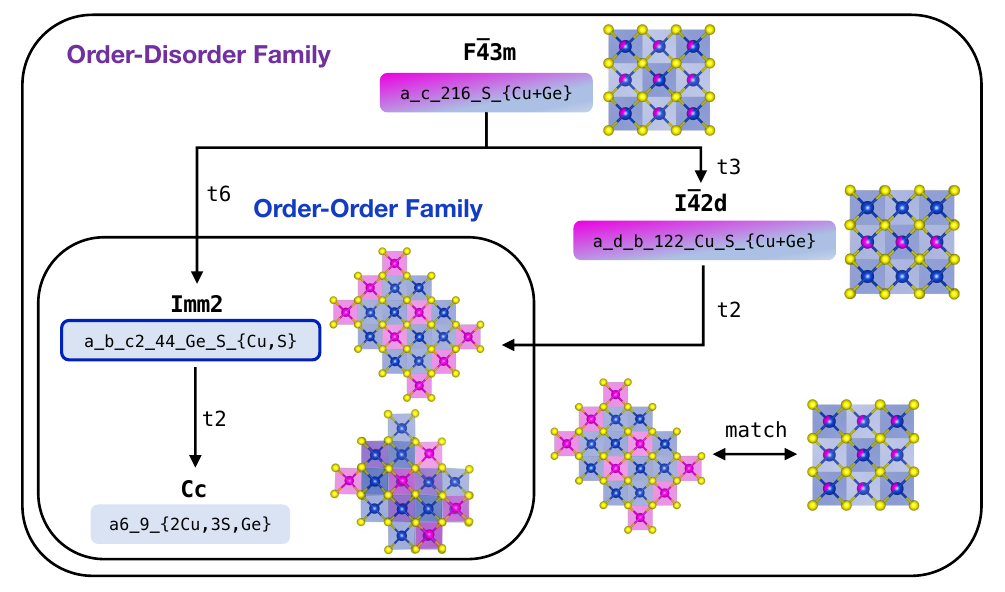}
    \caption{Order–(dis)order family tree recovered from the query structure Cu$_2$GeS$_3$  (\texttt{Imm2}, highlighted in bright blue). Each phase is labeled by its space group symbol and SWORD label, with parent-child links annotated by the subgroup index. The vertical arrangement follows point-group order, with higher-symmetry phases placed near the top. In this example, one of the disordered phases \texttt{a\_c\_216\_S\_\{Cu+Ge\}} occupies the highest-symmetry position and acts as the parent from which both ordered and disordered descendants are derived through group-subgroup transitions. Ordered children that share the same parent define an order-order subfamily within the broader order-(dis)order family tree. Two structures \texttt{match} if they share the same parent. Implementation details are provided in Appendix~\ref{sec:appendix_family_tree}.}
    \label{fig:fig1}
\end{figure}

Systematic navigation of order–disorder family trees requires a crystal representation that is both physically meaningful and computationally tractable. Our framework builds on two components: (i) the symmetry hierarchy of crystals expressed through space groups, Wyckoff positions, and group-subgroup transitions~\cite{Barnighausen1980, Wyckoff1922}; and (ii) a unified crystal representation describing both ordered and disordered structures within the same formal language. Together, these enable family-level reasoning where ordered children, disordered parents, and symmetry-related siblings are analyzed within a common framework.

\subsection{Preliminaries}
\paragraph{Space groups and Wyckoff positions}
The space group $G$ constitutes an important aspect of a crystal structure, representing the set of symmetry operations that leave the lattice invariant. Each operation $g \in G$ acts on a point $\mathbf{r} \in \mathbb{R}^3$ via an affine transformation $\mathbf{r} \mapsto R\mathbf{r} + \mathbf{t}$, where $R$ denotes a rotation or rotoinversion and $\mathbf{t}$ represents a translation. This group structure dictates the symmetry-equivalence of points within the crystal. For any specific point $\mathbf{r}$, the site-symmetry group is defined as the subgroup $G_{\mathbf{r}} = \{ g \in G \mid g\mathbf{r} = \mathbf{r} \}$, containing all operations that leave $\mathbf{r}$ fixed. Points are formally categorized into Wyckoff positions, which are sets of symmetry-equivalent sites whose site-symmetry groups are conjugate within $G$~\cite{Wyckoff1922}. Each Wyckoff position is uniquely characterized by its multiplicity, site symmetry, and fractional coordinates. Ultimately, Wyckoff positions provide a mathematically rigorous and compact description of crystal structures by grouping atomic sites according to their transformation properties under $G$.

\paragraph{Group-subgroup transitions}
Crystallographic relationships between structures are established using group--subgroup relations~\cite{Barnighausen1980}. For two space groups $G$ and $H$, the relation $H \leq G$ denotes that $H$ is a subgroup of $G$, such that all symmetry operations in $H$ are contained within $G$. A transition from $G$ to $H$ therefore corresponds to a reduction in symmetry. The degree of symmetry reduction is quantified by the subgroup index $[G : H] = |G| / |H|$, which counts how many cosets of $H$ partition $G$. A particularly important relation to consider is the \textit{translationengleiche} ($t$-type) subgroup relations, which preserve the translation group while lowering the point symmetry. These transitions are particularly relevant for tracing ordering processes in which previously equivalent sites become symmetry-distinct. Because $t$-type relations preserve the translational periodicity, this site-splitting accommodates ordering within the bounds of the original primitive cell, without introducing a supercell. Although \textit{klassengleiche} ($k$-type) subgroup relations, which involve a loss of translational symmetry, can also be important, they are not the main focus here. Under a group--subgroup transition, Wyckoff positions may split, causing previously equivalent sites to become distinguishable. This splitting provides the crystallographic basis for linking parent, child, and sibling structures within the same family tree.

\paragraph{Disorder}
Experimental structures often deviate from an ideal fully ordered configuration and may exhibit disorder in either site occupancy or atomic position~\cite{Antypov2025}. A broad and practically important class is \textit{occupational disorder}, in which the occupation of a crystallographic site is statistical rather than uniquely assigned. This includes cases where multiple chemical species share the same site with fractional occupancies, as well as vacancy-containing cases where a site is only partially occupied. Another major form is \textit{positional disorder}, in which atoms are distributed over multiple nearby positions instead of occupying a single well-defined site in the average structure. In this work, we focus primarily on occupational disorder and do not consider positional disorder, which constitutes less than 10\% of ICSD entries~\cite{Huang2026}.

\subsection{SWORD Representation}
Here, we use Symmetry and Wyckoff-sequence of Ordered and Disordered crystals (SWORD)~\cite{Huang2026}, a unified and scalable representation where both ordered and disordered crystal structures are encoded.  This enables consistent evaluation of family relations through space group symmetry-based analysis. SWORD expresses a crystal structure through symmetry-aware site assignments and occupancy information, making it particularly well suited for tracking how ordering, partial ordering, and disorder are distributed across Wyckoff sites. A SWORD label is written in the order of Wyckoff positions, space group number, and the atomic species occupying those positions. Using the disordered parent at the root of Figure \ref{fig:fig1} as an example, \texttt{a\_c\_216\_S\_\{Cu+Ge\}} denotes a structure in space group 216 in which Wyckoff position \texttt{WP a} is fully occupied by \texttt{Element S}, while Wyckoff position \texttt{WP c} is compositionally disordered between \texttt{Element Cu} and \texttt{Element Ge}. Likewise, one of its ordered children, \texttt{a\_b\_c2\_44\_Ge\_S\_\{Cu, S\}}, denotes a structure in space group 44 in which \texttt{WP a} is occupied by \texttt{Element Ge}, \texttt{WP b} by \texttt{Element S}, and the two distinct crystallographic orbits of \texttt{WP c} are occupied by \texttt{Element Cu} and \texttt{Element S}, respectively. This form of description is essential in our setting because the object of interest is not an isolated ordered phase, but the structural relation between an ordered derivative and its disordered or ordered relatives. By adopting a unified representation for both ordered and disordered phases, they can be compared at the level where order-disorder relationships are actually manifested.

\subsection{Order-(Dis)Order Family Trees Framework}
\label{sec:family_trees}
Family trees are constructed to recover, for a given structure (query), the set of crystallographically related phases connected by group-subgroup transitions and occupancy splitting, along which stoichiometry may vary. Starting from the query structure, candidate parent descriptions are systematically generated, and experimentally reported parent phases are subsequently identified using our \texttt{SWORDFamilyMatcher} module.

Figure \ref{fig:fig1} illustrates the concept of order--(dis)order family mapping using the family tree recovered from an ordered Cu$_2$GeS$_3$ query structure (\texttt{Imm2}, highlighted in bright blue) as an example. Each phase in this particular family tree corresponds to an experimentally reported ICSD entry and is labeled by its space group symbol and SWORD representation, while parent–child relations are defined by group–subgroup transitions and annotated with the corresponding subgroup index. The vertical arrangement follows point-group order, so that higher-symmetry phases appear higher in the tree. In this example, the disordered phase \texttt{a\_c\_216\_S\_\{Cu+Ge\}} occupies the highest-symmetry position and acts as the parent from which both ordered and disordered descendants are reached through group-subgroup transitions. From Figure \ref{fig:fig1}, it is clear that apparently distinct structures cannot be treated as isolated phases: an ordered phase may be placed within a broader family containing higher-symmetry disordered parents and symmetry-related ordered siblings. Our framework turns this intuition into a symmetry-guided mapping problem, in which a query structure is assigned to a crystallographic lineage rather than evaluated only as an isolated structure. The full connected set defines an order--(dis)order family tree, while ordered descendants sharing the same parent form order--order subfamilies. The same family tree would be recovered regardless of which phase in the tree is used as the starting query. The query therefore serves only as an entry point to the family. Additional implementation details are provided in Appendix~\ref{sec:appendix_family_tree}.

This matching framework captures several cases within a single rule: a predicted ordered structure may match an experimentally synthesized disordered parent, and two or more ordered structures may be linked through a shared parent, such that structures that are not identical at the child level are still recognized as members of the same broader family. In this way, the mapping lifts the structure matching from the level of isolated crystals to the level of symmetry-related families. Rather than asking only whether a predicted structure has been seen before in exactly the same ordered or disordered form, we also ask whether it already resides within an existing order-(dis)order lineage.

\setcounter{footnote}{0}
\begin{table}[h]
\centering
\caption{Comparison between ICSD Collection Codes of the previously reported disordered parents and those assigned by our framework for 22 A-Lab target phases. White rows (\checkmark) denote agreement with previously reported parent assignments. Grey rows ($\square$) indicate positional-disorder cases beyond the scope of the present framework. Green rows ($\ast$) highlight newly identified disordered parents, demonstrating that our framework both recovers known parents and identifies previously unreported ones.}
\label{tab:alab_comp}

\footnotesize 
\setlength{\tabcolsep}{5pt} 
\renewcommand{\arraystretch}{1.1} 

\begin{tabular}{c l l l l l}
\hline
\# & Target Phase & MP ID & ICSD Code\tablefootnote{Leeman et al.~\cite{Leeman2024}} & ICSD Code\tablefootnote{This work} & Remark\\
\hline
1 & Ba$_2$ZrSnO$_6$ & mp-1228067 & [43137] & [43137, 176330, 176331, 176329] & $\checkmark$ \\
2 & FeSb$_3$Pb$_4$O$_{13}$ & mp-1224890 & [60805] & [60805] & $\checkmark$ \\
3 & Hf$_2$Sb$_2$Pb$_4$O$_{13}$ & mp-1224490 & [62723] & [62723] & $\checkmark$ \\
4 & InSb$_3$Pb$_4$O$_{13}$ & mp-1223746 & [41119] & [41119] & $\checkmark$ \\
\rowcolor{gray!15}
5 & KMn$_3$O$_6$ & mp-1016190 & [240249] & NaN & $\square$ \\
6 & KNaP$_6$(PbO$_3$)$_8$ & mp-1223429 & [182501] & [182500, 182501, 182502] & $\checkmark$ \\
\rowcolor{gray!15}
7 & KNaTi$_2$(PO$_5$)$_2$ & mp-1211611 & [59284] & [67539, 71239] & $\square$ \\
\rowcolor{green!15}
8 & K$_2$TiCr(PO$_4$)$_3$ & mp-1224541 & NaN & [280999] & $\ast$ \\
\rowcolor{green!15}
9 & KPr$_9$(Si$_3$O$_{13}$)$_2$ & mp-1223421 & NaN & [153272] & $\ast$ \\
10 & K$_4$MgFe$_3$(PO$_4$)$_5$ & mp-532755 & [161484] & [161484] & $\checkmark$ \\
11 & K$_4$TiSn$_3$(PO$_5$)$_4$ & mp-1224290 & [250088] & [250088, 72720, 91534, 250087] & $\checkmark$ \\
\rowcolor{gray!15}
12 & NaCaMgFe(SiO$_3$)$_4$ & mp-1221075 & [75294] & [263074, 263075, 263076, 263077, 263078, $\dots$] & $\square$ \\
13 & Mg$_3$MnNi$_3$O$_8$ & mp-1222170 & [80306] & [80302, 80303, 80304, 80305, 80306, 80307] & $\checkmark$ \\
14 & Mg$_3$NiO$_4$ & mp-1099253 & [290603] & [290603, 13774] & $\checkmark$ \\
15 & MgTi$_2$NiO$_6$ & mp-1221952 & [171583] & [171583] & $\checkmark$ \\
16 & MgTi$_4$(PO$_4$)$_6$ & mp-1222070 & [74287] & [74287, 290277] & $\checkmark$ \\
17 & MgV$_4$Cu$_3$O$_{14}$ & mp-1222158 & [69731] & [69731, 69732] & $\checkmark$ \\
\rowcolor{green!15}
18 & Mn$_2$VPO$_7$ & mp-1210613 & NaN & [250126] & $\ast$ \\
19 & Mn$_4$Zn$_3$(NiO$_6$)$_2$ & mp-1222033 & [92223] & [92222, 92223, 92224] & $\checkmark$ \\
20 & Na$_3$Ca$_{18}$Fe(PO$_4$)$_{14}$ & mp-725491 & [85103] & [85103] & $\checkmark$ \\
21 & Sn$_2$Sb$_2$Pb$_4$O$_{13}$ & mp-1219056 & [62722] & [62722] & $\checkmark$ \\
22 & Zr$_2$Sb$_2$Pb$_4$O$_{13}$ & mp-1215826 & [62721] & [62721] & $\checkmark$ \\
\hline
\end{tabular}
\end{table}

\section{Results}

\subsection{Validation on A-Lab}
We first validate that our framework can recover disordered parents for experimentally synthesized structures. Applying it to the 35 GNoME phases reported as "successfully" synthesized by A-Lab~\cite{Szymanski2023}, we find 22 have experimentally reported disordered parent phases, either recovered by our framework or manually identified by Leeman et al~\cite{Leeman2024}. Table~\ref{tab:alab_comp} compares the ICSD Collection Codes of these previously reported disordered parents against those assigned by our framework for all 22 cases.

For 16 of the 22 (white-shaded $\checkmark$ rows), our framework systematically recovers the same disordered parents as Leeman et al.~\cite{Leeman2024}, directly validating the approach. Grey-shaded $\square$ rows involve positional disorder, which lies outside the current scope since our method targets occupational disorder; nevertheless, even in several of these cases our framework recovers alternative substitutionally disordered parents not previously reported.

\begin{figure}[t]
    \centering
    \includegraphics[width=\linewidth]{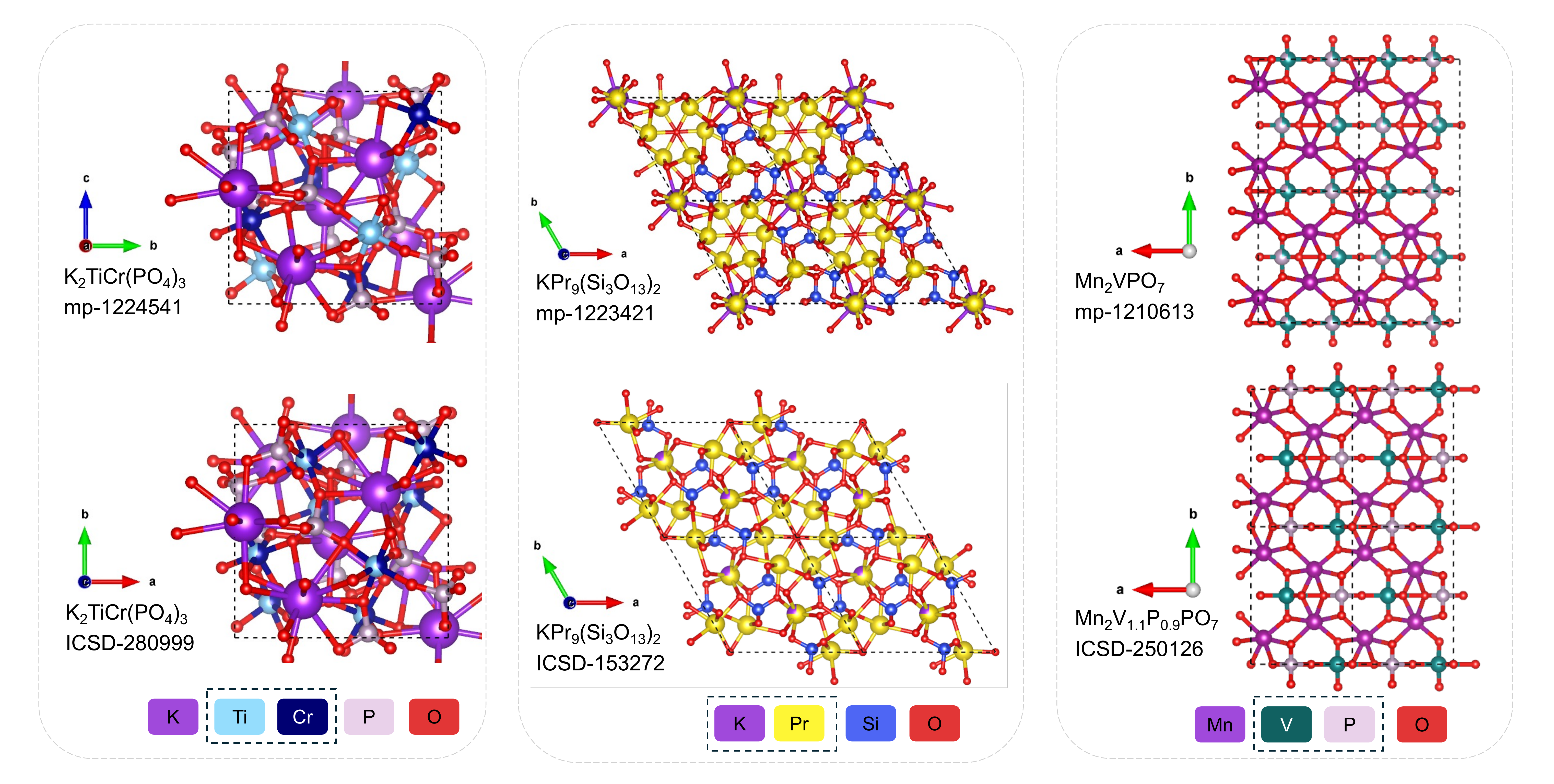}
    \caption{Recovery of experimentally reported disordered parents for three GNoME A-Lab target structures. Ordered GNoME structures are compared with the corresponding disordered parent phases reported in ICSD and recovered by our framework for \ce{K2TiCr(PO4)3}, \ce{KPr9(Si3O13)2}, and \ce{Mn2VPO7}. The colored labels at the bottom denote the elemental species, and elements enclosed by dashed outlines indicate disordered species.}
    \label{fig:fig2}
\end{figure}

Most notably, for the three green-shaded $\ast$ phases (\ce{K2TiCr(PO4)3}, \ce{KPr9(Si3O13)2}, \ce{Mn2VPO7}), prior analyses by Leeman et al.~\cite{Leeman2024} suggested the possible existence of disordered parents, but did not identify any explicitly. In contrast, our framework recovers corresponding disordered parents for all three cases. Figure \ref{fig:fig2} compares the target GNoME A-Lab ordered structures with their corresponding experimentally reported disordered parent phases in ICSD for these cases. Together, these results demonstrate the framework's capacity to systematically navigate order–disorder family trees, including cases that are difficult to identify through manual inspection alone.

\subsection{Order-(Dis)Order Family Tree Benchmark}
Having validated the framework, we next use it as a benchmarking tool for family-level novelty assessment of ordered structures relative to experimental order–disorder family trees in ICSD. Rather than evaluating a structure set only through isolated child-level matches, we assess it through the family relations recovered by the proposed matching framework. To this end, we introduce two complementary Family Tree based metrics, $FT_{\text{disorder}}$ and $FT_{\text{order}}$, that quantify how a given set of ordered structures is situated with respect to existing family trees.  $FT_{\text{disorder}}$ measures the fraction of ordered structures that fall within family trees rooted in existing disordered ICSD parents. A high value of $FT_{\text{disorder}}$ indicates that many ordered structures can be traced to known disordered phases, suggesting that their apparent ordering could be artificial and that experimentally realized compounds instead favor occupational disorder. As an example, in the set of 35 GNoME A-Lab structures, 21 are found by our framework to have existing disordered parents, corresponding to $FT_{\text{disorder}} = 60\%$. Meanwhile, $FT_{\text{order}}$ measures the fraction of ordered query structures that fall within family trees already spanned by existing ordered ICSD structures, i.e., those sharing the same disordered parent as their root, whether or not such a parent is experimentally synthesized. Unlike $FT_{\text{disorder}}$, a high $FT_{\text{order}}$ does not imply experimental suspicion; rather, it quantifies how much of the set of ordered structures lies within the known ordered family manifold formalized by group--subgroup relations, that is, structures that can in principle be derived from already synthesized ordered phases by rearranging atoms over specific crystallographic sites within a common host structural network, with accompanying symmetry lowering or raising. Conversely, a low $FT_{\text{order}}$ suggests genuinely new family trees absent from experimentally known crystallographic records.

\begin{table}[t]
\centering
\caption{Order-(dis)order family tree benchmark on known databases. For each dataset, $FT_{\text{disorder}}$ represents the percentage of ordered structures that fall within order-(dis)order families containing an existing disordered parent, and $FT_{\text{order}}$ represents the percentage that fall within families already spanned by existing ordered structures through group-subgroup relations. $n_{\text{disorder}}^{FT}$ and $n_{\text{order}}^{FT}$ are the corresponding numbers of structures assigned to these two categories, and $n_{\text{total}}$ is the total number of ordered structures evaluated.}
\label{tab:fam_bench_data}

\setlength{\tabcolsep}{8pt}
\renewcommand{\arraystretch}{1.2}

\begin{tabular}{lccccc}
\hline
Dataset & $FT_{\text{disorder}}$ (\%) & $FT_{\text{order}}$ (\%) 
& $n_{\text{disorder}}^{FT}$ & $n_{\text{order}}^{FT}$ 
& $n_{\text{total}}$ \\
\hline
$\mathrm{GNoME}_{\text{A-Lab}}$ & 60.00 & 0.00 & 21 & 0 & 35 \\
$\mathrm{ICSD}_{\text{order}}$ & 6.13 & 10.37 & 3,564  & 6,026 & 58,116 \\
$\mathrm{MP\text{-}20}_{\text{train+val}}$ & 23.27 & 11.16 & 8421 & 4,039 & 36,183 \\

$\mathrm{Alex\text{-}MP\text{-}20}_{\text{train+val}}$ & 3.75 & 2.39 & 1,876 & 1,196 & 50,000 \\

$\mathrm{GNoME}_{\text{stable}}$ & 0.73 & 0.40 & 367 & 199 & 50,000 \\

\hline
\end{tabular}
\end{table}

\paragraph{Benchmark on Databases.}
We evaluate $FT_{\mathrm{disorder}}$ and $FT_{\mathrm{order}}$ across both experimental and computational structure databases, including ICSD, MP-20, Alex-MP-20, and GNoME. Table~\ref{tab:fam_bench_data} summarizes these statistics alongside the corresponding numbers of structures assigned to each category.

Ordered structures in ICSD, the largest experimental crystal structure database to date, exhibit $FT_{\text{disorder}} = 6.13\%$. This reveals a nontrivial regime in which a single underlying structural network has been realized in both ordered and disordered forms, connected through group–subgroup relations. One possible explanation is that a subset of these ordered structures reflects incomplete structural refinement, where compositional disorder was not fully resolved or reported~\cite{Spek2018}. More plausibly, however, these cases represent genuine order-disorder transitions within a group-subgroup crystallographic lineage~\cite{Gratias2020}. In either case, the result highlights that even within the experimental record, a fraction of ordered compounds are not isolated from disorder-rooted families. A more pronounced effect is observed in MP-20~\cite{Xie2021}, a legacy Materials Project subset derived from ICSD structures with at most 20 atoms per unit cell. Here, $FT_{\mathrm{disorder}}$ increases sharply to $23.27\%$, indicating that a substantial fraction of structures coincide with experimentally realized disordered parent phases. We defer a more detailed analysis of this behavior to the Appendix~\ref{sec:appendix_mp_20}.

By contrast, computational structure databases exhibit substantially lower $FT_{\mathrm{disorder}}$ values. For 50k subsamples of Alex-MP-20~\cite{Zeni2025} and GNoME stable structures on the MP hull~\cite{Merchant2023}, we obtain $FT_{\mathrm{disorder}} = 3.75\%$ and $0.73\%$, respectively. This reduction should not be interpreted as evidence that computational structures are intrinsically less prone to disorder. Rather, $FT_{\mathrm{disorder}}$ is best understood as a lower bound on disorder propensity: it only counts cases for which a potential disordered parent phase can be found in existing experimental references. Experimental databases are concentrated in well-explored chemical systems, where both ordered and disordered phases have often been identified and recorded. Computational datasets, in contrast, extend into sparsely explored regions of composition space. In these regions, an ordered structure could still belong to a broader disorder-rooted family, but the corresponding disordered parent might simply be absent from current databases. As a result, the lower $FT_{\mathrm{disorder}}$ observed for computational datasets more likely reflects incomplete coverage of disorder families in experimental references than a true absence of disorder. Consistent with this view, manual inspection of GNoME structures reveals that many candidates can be associated with plausible disordered parents in ICSD~\cite{Cheetham2024}; moreover, recent predictive studies also suggest that a majority of GNoME structures are in fact prone to occupational disorder~\cite{Jakob2025}.

A complementary perspective emerges from $FT_{\mathrm{order}}$. In ICSD, more than $10\%$ of ordered structures belong to order-order families, indicating that a noticeable fraction of experimentally reported phases within the same composition space are related through group-subgroup transitions on a common structural network. MP-20 shows a similar level of overlap, whereas Alex-MP-20 and GNoME exhibit lower fractions, potentially because their broader compositional sampling reduces the likelihood that multiple ordered members of the same family are simultaneously represented. At the same time, the lower values of $FT_{\mathrm{order}}$, which explicitly account for chemical identity, may indicate that although a structure is new for a particular combination of elements, its underlying structure type has already been explored elsewhere~\cite{Eckert2024}.

\begin{table}[t]
\centering
\caption{Order-(dis)order family tree benchmark on 10{,}000 samples from crystal generative models trained on MP-20 and Alex-MP-20. For each model, $FT_{\text{disorder}}$ represents the percentage of ordered structures that fall within order-(dis)order families containing an existing disordered parent, and $FT_{\text{order}}$ represents the percentage that fall within families already spanned by existing ordered structures through group-subgroup relations. Lower values indicate weaker overlap with experimentally known family manifolds and therefore greater family-tree-based novelty.}
\label{tab:fam_bench_gen}

\setlength{\tabcolsep}{7pt}
\renewcommand{\arraystretch}{1.2}

\begin{tabular}{l l c c}
\hline
Dataset & Model & $FT_{\text{disorder}} \downarrow$ (\%) & $FT_{\text{order}} \downarrow$ (\%) \\
\hline
MP-20      & DiffCSP     & 4.88    & 10.13 \\
           & DiffCSP++   & 4.48    & 4.44  \\
           & FlowMM      & 4.23    & 7.79  \\
           & MiAD        & 14.24   & 25.43 \\
           & SymmCD      & 3.83    & \textbf{3.26}  \\
           & CrystalDiT  & 5.05    & 13.69 \\
           & ADiT        & 12.61   & 28.12 \\
           & MatterGen   & 6.88    & 10.11 \\
           & Chemeleon2-RL   & 6.65    & 6.04 \\
           & WyFormer    & \textbf{3.40} & 4.00 \\
\hline
Alex-MP-20 & MatterGen   & 3.67    & 3.01 \\
           & WyFormer    & \textbf{0.96} & \textbf{1.17} \\
\hline
\end{tabular}
\end{table}

\paragraph{Benchmark on Generative Models.}
We next benchmark two major classes of state-of-the-art crystal generative models trained on MP-20: (i) all-atom diffusion or flow-matching models and (ii) space group symmetry-based models, spanning 10 models in total. Category (i) includes DiffCSP, FlowMM, MiAD, CrystalDiT, ADiT, Chemeleon2-RL, and MatterGen, whereas category (ii) includes DiffCSP++, SymmCD, and WyFormer. Their central difference lies in the space over which generation is parameterized. All-atom models operate directly on the full set of atoms in the unit cell, without explicitly restricting the generative trajectory by crystallographic symmetry. Symmetry-based models instead operate in the asymmetric-unit, or Wyckoff subspace, representing crystals through discrete symmetry elements such as space group and site symmetry within a constrained parameterization. This both regularizes the search space and reduces the effective number of degrees of freedom. The two classes therefore offer a natural comparison between symmetry-agnostic generation and generation that is symmetry-aware by design.

Table~\ref{tab:fam_bench_gen} reports the order-(dis)order family tree benchmark for the 10 generative models trained on MP-20, together with MatterGen and WyFormer additionally trained on Alex-MP-20. The most immediate observation is the elevated $FT_{\mathrm{disorder}}$ of MiAD and ADiT, at $14.24\%$ and $12.61\%$, respectively, followed by MatterGen. Notably, all three belong to the symmetry-agnostic all-atom models. These high values indicate that a substantial fraction of generated structures that appear novel at the ordered-child level may in fact fall within family trees whose disordered parent phases are already known experimentally. In other words, apparent novelty at the ordered level can mask a closer relationship to previously realized disorder-rooted families. This trend is particularly notable in light of the reported synthesis case of $\mathrm{TaCr_2O_6}$ generated by MatterGen~\cite{Juelsholt2026}. Although initially proposed as a new ordered structure, the synthesized product was later found to correspond instead to a Cr-rich variant of an existing disordered parent phase. In this context, the high $FT_{\mathrm{disorder}}$ values of this model class suggest that the reported example may be symptomatic of a more general trend.

By contrast, WyFormer achieves the lowest $FT_{\mathrm{disorder}}$, at $3.40\%$, followed by SymmCD, and both belong to the space group symmetry-based models. A similar pattern is observed when MatterGen and WyFormer are trained on Alex-MP-20: WyFormer again exhibits an $FT_{\mathrm{disorder}}$ nearly four times lower than that of MatterGen. Together, these results suggest that symmetry-regularized models are substantially less likely to generate ordered structures that fall within family trees of known disordered parents, and are therefore less prone to producing ordered candidates shadowed by experimentally reported disordered phases. A similar trend is observed for $FT_{\mathrm{order}}$. All-atom models again show consistently higher values, indicating that their generations more frequently occupy regions of already known order-order family space. By contrast, the lower $FT_{\mathrm{order}}$ of symmetry-based models suggests that they are more likely to generate ordered structures that are novel not only as individual entries, but also at the family level.

This contrast is consistent with the design of the two model classes. WyFormer, one of the leading symmetry-constrained models, achieves the highest uniqueness and novelty in the Wyckoff subspace relative to symmetry-agnostic models such as ADiT and MatterGen~\cite{Betala2025}. Our results suggest that this advantage at the child level may translate into the family level: models designed to operate explicitly in a symmetry-constrained design space appear better positioned to discover genuinely new family-level configurations, rather than merely new ordered structures within already known families. This should be a consideration for future generative models in order to achieve the goal of synthesizing truly novel materials in experiments.

\paragraph{Stable P1.}
Space group \texttt{P1} is the simplest crystallographic symmetry, containing no symmetry operations beyond identity and lattice translations. In nature, such trivial symmetry is rare: its occurrence in ICSD amounts to less than 1\%, and even this fraction may be inflated by misreported cases~\cite{Urusov2009, Marsh1999}. In contrast, all-atom generative models are known to overproduce \texttt{P1} structures, which in some cases constitute nearly half of generated samples, far exceeding the experimental distribution. Whether these \textit{novel} and \textit{stable} \texttt{P1} structures represent a byproduct of imperfect computational evaluation or a genuinely unexplored class of experimentally accessible structure types remains an open question in the community.

\begin{figure}[t]
    \centering
    \includegraphics[width=\linewidth]{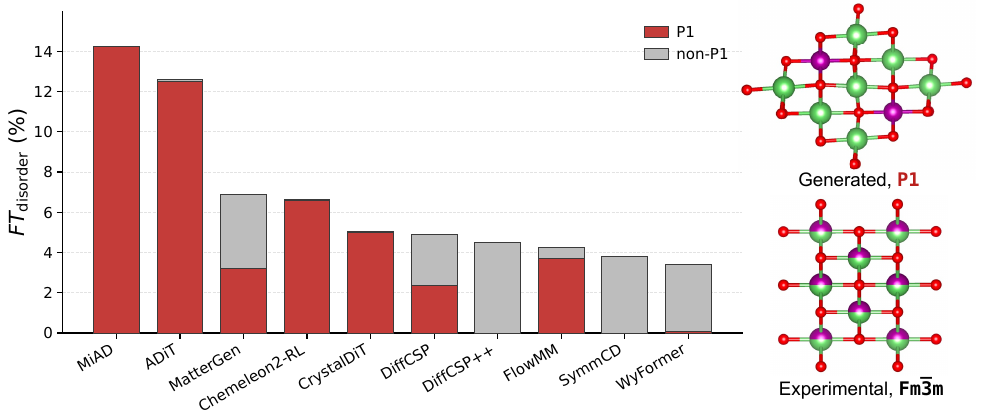}
    \caption{Space group decomposition of $FT_{\mathrm{disorder}}$ across crystal generative models. For each model, $FT_{\mathrm{disorder}}$ is decomposed by the space group of the ordered generated structure, with \texttt{P1} highlighted in red. The dominant contribution of \texttt{P1} in all-atom models points to the possibility that some of the overproduced P1 structures originate as ordered children of known disordered parent phases. Right, an example of generated \texttt{P1} child and its matched ICSD disordered parent in \texttt{Fm\={3}m}.}
    \label{fig:fig3}
\end{figure}

To examine this, we analyzed the space group distribution of ordered structures identified as children of existing disordered parents for each model. Figure~\ref{fig:fig3} highlights the $FT_{\mathrm{disorder}}$ of each model, with \texttt{P1} contributions marked in red. For all-atom models, including MiAD, ADiT, MatterGen, Chemeleon2-RL, CrystalDiT, DiffCSP, and FlowMM, \texttt{P1} dominates $FT_{\mathrm{disorder}}$, accounting for anywhere from nearly half to essentially all child structures, making it apparent that a part of \texttt{P1} structures systematically appear as ordered configurations derived from existing disordered parent phases. An example is shown on the right side of Figure~\ref{fig:fig3}, where a \texttt{P1} structure generated by DiffCSP is traced back to an ICSD disordered parent crystallizing in space group \texttt{Fm\={3}m}, using our framework. 

Further DFT validation of \texttt{P1} structures that have ICSD disordered parents, with MiAD and MatterGen taken as representative all-atom models, reveals that roughly 80–90\% are metastable after relaxation, with energies above the hull below 100 meV/atom relative to the Materials Project database~\cite{Horton2025}. Details of the DFT calculations are provided in Appendix~\ref{sec:appendix_dft}. These findings, therefore, suggest that the overproduction of apparently stable \texttt{P1} structures can be explained to some extent by their emergence as ordered children of high-symmetry disordered parent phases: configurations that, while compositionally disordered in experiment, are unlikely to be experimentally synthesized in their ordered \texttt{P1} form.

\paragraph{Group-Subgroup Polymorphs.}
Order-order family trees, as formalized in Section~\ref{sec:family_trees}, are governed by group-subgroup relations. In general, ordered structures belonging to the same $FT_{\text{order}}$ family do not necessarily share the same stoichiometry, because symmetry lowering can split Wyckoff orbits in ways that alter the allowed occupancy pattern. A particularly important subset arises when symmetry descent preserves the exact same stoichiometry. We refer to this subset as \textit{group–subgroup polymorphs}.

These polymorphs represent a symmetry-grounded form of polymorphism: one ordered phase can be derived from another through a group-subgroup transition that lowers point-group symmetry while preserving composition. This perspective goes beyond treating polymorphs simply as distinct structures with the same stoichiometry, and instead isolates those connected by an explicit symmetry-descent pathway. A representative example is provided by the two ordered Cu$_2$GeS$_3$ phases in Figure~\ref{fig:fig1}, experimentally reported in space groups \texttt{Imm2} and \texttt{Cc}, where one can be obtained from the other through symmetry breaking along a group-subgroup transition.

This behavior is not rare in experimental data. In ICSD, among roughly $5 \times 10^{4}$ unique ordered compositions, about $10\%$ exhibit polymorphism, and around $40\%$ of those polymorphic systems contain at least one pair related by a group-subgroup transition. This indicates that a meaningful fraction of experimentally reported polymorphs are not merely alternative structural realizations of the same composition, but members of an explicit symmetry-connected lineage, namely an order-order family. Order--order family mapping therefore offers more than a descriptive framework for known polymorphs, but it could also provide a systematic basis for discovering new ones. 

\section{Discussion}

The order-(dis)order family tree benchmark introduced here redefines novelty against experimentally grounded crystallographic lineage rather than isolated ordered endpoints. This family-level view provides a more meaningful notion of novelty in systems where multiple ordered phases descend from a common disordered parent. The benchmark is most informative when the relevant disordered parent is already present in the experimental record. For underexplored systems, the corresponding parent may simply remain unseen, and the absence of a family assignment marks regions of family space not yet experimentally charted rather than a genuine absence of disorder tendency. 

More fundamentally, these results indicate that disorder is best treated not as a label to predict, but as a manifold to navigate. Extending access to unseen disordered parents is therefore a natural next step: it would improve how novelty is assessed, clarify what synthesis is actually likely to realize, and provide a broader structural basis for evaluating stability across experimentally relevant energetic landscapes. In this sense, the same family-manifold perspective should ultimately inform not only crystallographic novelty, but also how thermodynamic preference is estimated within an order-disorder family~\cite{Divilov2024}. The order-(dis)order family tree framework provides a concrete structural basis for that shift.

\section{Conclusion}

Across A-Lab case studies, known structure databases (ICSD, MP-20, Alex-MP-20, GNoME), and state-of-the-art crystal generative models, we found that many ordered structures that appear novel under conventional child-level comparison are more naturally interpreted as members of already known family trees, often as ordered children of existing disordered parents. This effect is especially visible in synthesis-facing cases and in symmetry-agnostic all-atom generative models, where the labeled novelty can be traced to originate from rediscovery of ordered configurations of experimentally realized disordered parents. We further show that the excessive number of apparently novel and stable \texttt{P1} structures produced by all-atom models can, in part, be understood as ordered children of existing disordered parents. In parallel, the framework also exposes symmetry-connected order-order families and identifies a meaningful subset of experimentally reported polymorphs as group-subgroup polymorphs.

Our results further reveal that symmetry-constrained generative models provide a more promising route to genuine family-level discovery than symmetry-agnostic all-atom models. By operating directly in a symmetry-constrained design space, they are less likely to generate ordered structures that simply occupy already known order-(dis)order families, and are better positioned to explore new crystallographic lineage at the family level. 

In this work, we introduced order-(dis)order family trees as a symmetry-grounded framework for organizing ordered and disordered crystal structures through group-subgroup relations. By shifting the unit of analysis from isolated structures to crystallographic families, the framework makes it possible to assess novelty at the level most relevant to experimental realization.  We therefore argue that order-(dis)order family trees should be treated as a key requirement in evaluating novelty in materials discovery systems. Overall, these results position disorder-aware family matching as a practical foundation for future discovery pipelines and a necessary step toward identifying genuinely new, experimentally realizable materials.

\section*{Data Availability}
The crystal structure data used in this study were obtained from the Inorganic Crystal Structure Database (ICSD), version \texttt{2025.2\_v5.5.0}. ICSD entries are available at \url{https://icsd.products.fiz-karlsruhe.de} under a commercial license.

\section*{Code Availability}
Data curation of the ICSD prior to analysis was performed using code available at \url{https://github.com/YuyaoHuang330/SWORD}. The implementation of the order-(dis)order family tree framework is available at \url{https://github.com/shuyayamazaki/order-disorder-family-trees}.

\section*{Acknowledgments}
We thank Andrey Okhotin for providing data samples for DiffCSP, DiffCSP++, FlowMM, MiAD, SymmCD, and CrystalDiT. K.H. acknowledges funding from the MAT-GDT Program at A*STAR via the AME Programmatic Fund by the Agency for Science, Technology and Research under Grant No. M24N4b0034. M.H.P. acknowledges funding from the NTU AI4X Postdoctoral Fellowship.

\bibliographystyle{unsrt}  
\bibliography{refs}  

\clearpage
\appendix
\renewcommand{\thetable}{S\arabic{table}}
\renewcommand{\thefigure}{S\arabic{figure}}
\setcounter{table}{0}
\setcounter{figure}{0}

\section{Order-(Dis)Order Family Matching Framework}
\label{sec:appendix_family_tree}

Our order--(dis)order family matching framework is implemented in the \texttt{SWORDFamilyMatcher} module. For a query crystal structure $S$, the procedure first computes its SWORD child label,
\begin{equation*}
\ell_{\mathrm{child}}(S)=\mathrm{SWORD}(S),
\end{equation*}
using symmetry analysis with \texttt{spglib}~\cite{Togo2024} through the SWORD labeling routine. In practice, the structure is first parsed into a \texttt{pymatgen} \texttt{Structure}~\cite{Ong2013},  and partially occupied sites are renormalized when needed so that occupancies sum to unity before label construction. The overall procedure is summarized in Algorithm~\ref{alg:sword-family-matcher}.

The central idea is to infer possible higher-symmetry parent descriptions by selectively removing chemical distinctions and recomputing the corresponding symmetry-aware label. If an ordered or partially ordered structure becomes compatible with a more symmetric description once certain species distinctions are masked, then that higher-symmetry description is taken as a plausible parent candidate.

Let $E(S)$ denote the set of distinct element types in $S$, with $n=|E(S)|$. For a chosen subset $M \subseteq E(S)$, we define a masking operation $\mathcal{M}_M(S)$ that replaces every element in $M$ by a common dummy species $X$. Recomputing the SWORD label after masking gives a candidate parent label,
\begin{equation*}
\ell_{\mathrm{parent}}^{(M)}(S)=\mathrm{SWORD}\!\left(\mathcal{M}_M(S)\right).
\end{equation*}
Because the raw label obtained after masking is expressed in terms of the dummy species $X$, we then restore the chemically meaningful mixed-site description by replacing each occurrence of $X$ with the corresponding masked element set. This yields the final parent label used for matching.

For an ordered query, all nontrivial masks with $2 \leq |M| \leq n$ are enumerated, giving
\begin{equation*}
\sum_{k=2}^{n}\binom{n}{k}=2^n-n-1
\end{equation*}
possible masking patterns. For a query that is already disordered, masking is restricted to the mixed-element groups already present in the SWORD child label. This restriction is important because an element may appear simultaneously on ordered and disordered sites, so the relevant family relation is not obtained by arbitrary recombination of all species. Instead, the existing mixed-occupancy pattern encoded in the child label is treated as the chemically meaningful starting point, and only those disorder groups are further abstracted when generating parent candidates.

Accordingly, the parent-label set is written as
\begin{equation*}
\mathcal{P}(S)=\left\{ \ell_{\mathrm{parent}}^{(M)}(S)\right\},
\end{equation*}
and the family label set of the query is
\begin{equation*}
\mathcal{F}(S)=\left\{\ell_{\mathrm{child}}(S)\right\}\cup \mathcal{P}(S).
\end{equation*}
Given a reference structure $R$ processed in the same way, we regard $R$ as family-related to $S$ whenever their family label sets intersect,
\begin{equation*}
\mathcal{F}(S)\cap\mathcal{F}(R)\neq\varnothing.
\end{equation*}

For ordered queries, the implementation includes an additional vacancy-ordering augmentation step before parent enumeration. If the child label contains no mixed occupancy marker, the structure is analyzed for symmetry-compatible latent vacancy ordering using the \texttt{find\_vacancy\_ordered} routine. This routine identifies candidate void sites from a periodic Voronoi construction, evaluates their local environments against species-specific coordination templates, and fills only those sites whose geometry and nearest-neighbor spacing are consistent with the insertion of an existing species. When such a filled structure is obtained, parent labels are generated both from the original query and from the vacancy-filled variant. This step allows the matcher to recover family relations that may otherwise remain hidden when an experimentally relevant parent description is more naturally expressed after restoring an ordered vacancy configuration.

In the pairwise setting, two structures $S_1$ and $S_2$ are assigned to the same family if
\begin{equation*}
\mathcal{F}(S_1)\cap \mathcal{F}(S_2)\neq\varnothing.
\end{equation*}
In database matching, the same criterion is used against every reference entry. If the matched reference child label contains mixed occupancy, the hit is recorded as a disordered-family match; otherwise, if the matched reference child label is ordered and distinct from the query child label, it is recorded as an order--order family match.

\begin{algorithm}[H]
\caption{\textsc{SWORDFamilyMatcher}}
\label{alg:sword-family-matcher}
\begin{algorithmic}[1]
  \STATE \textbf{Input:} Query structure $S$ or two structures $S_1,S_2$
  \STATE \textbf{Output:} Family label set for $S$, or a Boolean family match for $S_1,S_2$

  \STATE
  \STATE \textbf{Function} \textsc{GetFamilyLabels}$(S)$
    \STATE parse $S$ into a crystal structure object
    \STATE renormalize partial occupancies if needed
    \STATE $\ell_{\mathrm{child}} \gets \mathrm{SWORD}(S)$
    \STATE $\mathcal{S}_{\mathrm{proc}} \gets [S]$

    \IF{$\ell_{\mathrm{child}}$ contains no mixed-occupancy marker}
      \STATE $S_{\mathrm{fill}} \gets \textsc{FindVacancyOrdered}(S)$
      \IF{$S_{\mathrm{fill}}$ is valid}
        \STATE append $S_{\mathrm{fill}}$ to $\mathcal{S}_{\mathrm{proc}}$
      \ENDIF
    \ENDIF

    \STATE $\mathcal{P} \gets \varnothing$
    \FOR{each $\widetilde{S} \in \mathcal{S}_{\mathrm{proc}}$}
      \STATE $\widetilde{\ell}_{\mathrm{child}} \gets \mathrm{SWORD}(\widetilde{S})$
      \STATE extract mixed-element groups from $\widetilde{\ell}_{\mathrm{child}}$

      \IF{mixed-element groups are present}
        \STATE define mask sets from those existing mixed groups only
      \ELSE
        \STATE define all element subsets $M \subseteq E(\widetilde{S})$ with $2 \le |M| \le |E(\widetilde{S})|$
      \ENDIF

      \FOR{each mask set $M$}
        \STATE construct $\mathcal{M}_M(\widetilde{S})$ by replacing elements in $M$ with dummy species $X$
        \STATE $\ell_{\mathrm{raw}} \gets \mathrm{SWORD}(\mathcal{M}_M(\widetilde{S}))$
        \STATE $\ell_{\mathrm{parent}} \gets$ restore masked element group into $\ell_{\mathrm{raw}}$
        \STATE add $\ell_{\mathrm{parent}}$ to $\mathcal{P}$
      \ENDFOR
    \ENDFOR

    \RETURN $\mathcal{F}(S)=\{\ell_{\mathrm{child}}\}\cup\mathcal{P}$

  \STATE
  \STATE \textbf{Function} \textsc{Fit}$(S_1,S_2)$
    \STATE $\mathcal{F}_1 \gets \textsc{GetFamilyLabels}(S_1)$
    \STATE $\mathcal{F}_2 \gets \textsc{GetFamilyLabels}(S_2)$
    \IF{$\mathcal{F}_1 \cap \mathcal{F}_2 \neq \varnothing$}
      \RETURN \texttt{True}
    \ELSE
      \RETURN \texttt{False}
    \ENDIF
\end{algorithmic}
\end{algorithm}

As a concrete example, for the ordered Cu$_2$GeS$_3$ query structure used in Figure~\ref{fig:fig1}, \texttt{SWORDFamilyMatcher} returns the following SWORD family dictionary:
\begin{verbatim}
{'child_label': 'a_b_c2_44_Ge_S_{Cu,S}',
 'parent_labels': ['a_c_216_S_{Cu+Ge}',
  'a_b_c2_44_Ge_{Cu+S}_{2(Cu+S)}',
  'a_b_c2_44_{Ge+S}_{Ge+S}_{Cu,Ge+S}',
  'a_227_{Cu+Ge+S}'],
 'matched_ordered_labels': ['a6_9_{2Cu,3S,Ge}'],
 'matched_ordered_ids': [85138, 146583],
 'matched_disordered_labels': ['a_d_b_122_Cu_S_{Cu+Ge}', 'a_c_216_S_{Cu+Ge}'],
 'matched_disordered_ids': [627781, 43531, 102963, 100955, 627773]}
\end{verbatim}
This concrete example shows how the matcher encodes the query structure, its candidate parent descriptions, and the ordered and disordered ICSD relatives that together define the family recovered in Figure~\ref{fig:fig1}.

\section{MP-20}
\label{sec:appendix_mp_20}

MP-20 has long been treated as one of the most representative training datasets for crystal generative models. It is defined as an ICSD-derived subset of Materials Project structures restricted to compounds with at most 20 atoms per unit cell, and contains roughly $45\,\mathrm{k}$ structures across the train, validation, and test splits. At the same time, when ordered ICSD entries are counted at the level of unique structures with no more than 20 atoms per unit cell, the total is only around $20,\mathrm{k}$. This discrepancy is difficult to reconcile with a strict one-to-one view of MP-20 as a subset of experimentally reported ordered ICSD phases.

One plausible explanation is that the ICSD provenance inherited in \textit{legacy} Materials Project data is not always cleanly recoverable at the level of unique current ICSD entries, and that MP-20 also includes Materials Project processed or relaxed versions of experimentally derived structures. Under this interpretation, MP-20 should be understood not as a strict snapshot of unique ordered ICSD phases, but as a legacy computational dataset with partial ICSD provenance and additional expansion introduced during downstream curation.

This distinction matters for interpreting its family-tree statistics. In particular, the fact that $FT_{\mathrm{disorder}}$ reaches $23.27\%$ indicates that more than one in five ordered structures in MP-20 have existing disordered parents in ICSD. This suggests that a non-negligible portion of MP-20 lies in crystallographic families for which disorder is already an experimentally established realization. As a consequence, models trained on MP-20, particularly all-atom generative models, may in part be learning to reproduce ordered children of existing disordered parents rather than structures that are genuinely novel at the family level. At a high level, this observation further motivates disorder-aware dataset construction and generative modeling in crystal structure generation.

\section{DFT details}
\label{sec:appendix_dft}
We use DFT settings from Materials Project \url{https://docs.materialsproject.org/methodology/materials-methodology/calculation-details/gga+u-calculations/parameters-and-convergence} for structure relaxation and energy computation. In particular, we do GGA and GGA+U calculations with \texttt{atomate2.vasp.flows.mp. MPGGADoubleRelaxStaticMaker}~\citep{Ganose2025}, which in turn relies on \texttt{pymatgen.io.vasp.sets.MPRelaxSet} and \texttt{pymatgen.io.vasp.sets.MPStaticSet}~\citep{Ong2013}. Computations themselves were done with VASP~\citep{Kresse1996} version 5.4.4. with the plane-wave basis set~\citep{Kresse1996}. The electron-ion interaction is described by the projector augmented wave (PAW) pseudo-potentials~\citep{Kresse1999}. The exchange-correlation of valence electrons is treated with the Perdew-Burke-Ernzerhof (PBE) functional within the generalized gradient approximation (GGA)~\citep{Perdew1996}. The raw total energies computed by DFT were corrected with \texttt{MaterialsProject2020Compatibility} before putting into the \texttt{PhaseDiagram} to obtain the DFT $E_\text{hull}$. We used the MP convex hull \texttt{2023-02-07-ppd-mp.pkl.gz} distributed by \texttt{matbench-discovery}~\citep{Riebesell2025} as the reference hull.

\end{document}